\documentclass[aps,prl,reprint,superscriptaddress,dvipdfm]{revtex4-1}

\usepackage[pdftex]{graphicx}
\usepackage{longtable}
\usepackage{amsmath,amssymb}

\usepackage{physics}
\usepackage{mathtools}

\begin{document}

\title{
Hyperfine-structure-resolved laser spectroscopy of many-electron highly charged ions
}

\author{Naoki Kimura}\email[]{Corresponding author: naoki.kimura@riken.jp}
\affiliation{Atomic, Molecular and Optical Physics Laboratory, RIKEN, Saitama 351-0198, Japan}
\author{Priti}\email[]{Present address: National Institute for Fusion Science, Toki, Gifu 509-5292, Japan}
\affiliation{Institute for Laser Science, The University of Electro-Communications, Tokyo 182-8585, Japan}
\author{Yasutaka Kono}
\affiliation{Institute for Laser Science, The University of Electro-Communications, Tokyo 182-8585, Japan}
\author{Pativate Pipatpakorn}
\affiliation{Institute for Laser Science, The University of Electro-Communications, Tokyo 182-8585, Japan}
\author{Keigo Soutome}
\affiliation{Institute for Laser Science, The University of Electro-Communications, Tokyo 182-8585, Japan}
\author{Naoki Numadate}\email[]{Present address: Komaba Institute for Science, The University of Tokyo, Tokyo 153-8902, Japan}
\affiliation{Institute for Laser Science, The University of Electro-Communications, Tokyo 182-8585, Japan}
\author{Susumu Kuma}
\affiliation{Atomic, Molecular and Optical Physics Laboratory, RIKEN, Saitama 351-0198, Japan}
\author{Toshiyuki Azuma}
\affiliation{Atomic, Molecular and Optical Physics Laboratory, RIKEN, Saitama 351-0198, Japan}
\author{Nobuyuki Nakamura}
\affiliation{Institute for Laser Science, The University of Electro-Communications, Tokyo 182-8585, Japan}

%\date{\today}

\pacs{}
\maketitle

\textbf{
Hyperfine-structures of highly charged ions (HCIs) are favourable spectroscopic targets for exploring fundamental physics as well as nuclear properties.
Recent proposals of HCI atomic clocks highlight their importance, especially for many-electron HCIs, and they have been theoretically investigated by refining atomic-structure calculations.
Nonetheless, no established spectroscopic method is currently available to verify these theoretical calculations.
Here, we demonstrate hyperfine-structure-resolved laser spectroscopy of HCIs in an electron beam ion trap plasma, employing the magnetic-dipole transition in 4$d^{9}$5$s$ of  $^{127}$I$^{7+}$. 
Ion-state manipulation by controlled electron collisions in the well-defined laboratory plasma enables laser-induced fluorescence spectroscopy of trapped HCIs.
The observed spectrum of evaporatively cooled ions under the low magnetic field shows remarkable features reflecting the hyperfine-structures.
The present demonstration using the combined optical and plasma approach provides a new benchmark for state-of-the-art atomic calculations of hyperfine-structures in many-electron HCIs and offers possibilities for a variety of unexploited experiments.
}

Nuclear-electron interactions induce particularly small splitting in atomic energy levels, defined as hyperfine-structure.
Spectroscopically investigating hyperfine-structures reveals highly valuable information related to nuclear properties and atomic structures \cite{Campbell2016}.
Extension of the spectroscopic target, from the usually studied neutral atoms or singly charged ions to highly charged ions (HCIs), represents an excellent approach to enhance hyperfine interactions owing to contracted electron clouds.
Spectroscopic measurements of hyperfine-structures in few-electron HCIs, such as H-, He-, Li-, and Be-like ions, have been widely performed by taking advantage of their enhanced energy intervals between the hyperfine levels.
They have successfully contributed to tests of relativistic and quantum electrodynamics (QED) atomic theories as well as investigations of nuclear properties \cite{Klaft1994, Seelig1998, Crespo1996, Crespo1998, Beiersdorfer2001, Ullmann2017, Beiersdorfer1998, Beiersdorfer2014, Myers1981,  Myers1994, Myers1996, Thompson1998, Myers1999-1}.

In contrast to few-electron HCIs, hyperfine-structure spectroscopy of many-electron HCIs provides distinctive examples to gain a deeper understanding about complex relativistic electron correlations.
Nevertheless, this implicates considerable experimental challenges as high spectral resolution (i.e., better than their slightly enhanced hyperfine-structure) are required.
Such HCIs have recently attracted much attention, triggered by proposals for their application in high-precision atomic clocks.
HCI clocks are expected to be a sensitive probe for time variations of the fine-structure constant $\alpha$ for testing modern physical theories beyond the Standard Model \cite{Kozlov2018}.
A number of candidates for the HCI clocks have been proposed \cite{Berengut2010, Berengut2011, Berengut2012-1, Berengut2012-2, Dzuba2012, Derevianko2012, Yudin2014, Safronova2014-1, Safronova2014-2, Safronova2014-3, Dzuba2015, Dzuba2016, Nandy2016, Cheung2020, Beloy2020, Yu2019}, and their transition wavelengths have been experimentally investigated \cite{Windberger2015, Bekker2019, Kimura2019, Liang2021}.
Micke $et.$ $al.$ recently demonstrated quantum logic spectroscopy of HCIs using Ar$^{13+}$, which has no hyperfine-structures, promising realization of the HCI clocks \cite{Micke2020, King2022}.
For further HCI atomic clock development employing fascinated transition targets, it is necessary to deepen the understanding of hyperfine-structures in many-electron HCIs.
The atomic-structure calculations have been developed to provide the hyperfine-structure constants \cite{Dzuba1984, Jonsson1993, Fischer1994, GRASP, AMBIT}, and several theoretical studies suggested that hyperfine interactions in promising HCI clock candidates substantially affect their clock operation and attainable uncertainties \cite{Berengut2010, Dzuba2015, Dzuba2016, Yu2019, Bekker2019}.
However, there have been no established spectroscopic methods to experimentally evaluate atomic calculations for the hyperfine-structures.

Here, we study the hyperfine-structures in the 4$d^{9}$5$s$ metastable states of Pd-like $^{127}$I$^{7+}$, using laser-induced fluorescence (LIF) spectroscopy for HCIs in an electron beam ion trap (EBIT) plasma.
The present method uses electric-dipole($E1$)-forbidden transitions both for laser excitation and fluorescence detection.
High-resolution spectroscopy of the transition between highly excited metastable states is achieved with the aid of excitation processes in the EBIT plasma.
A similar experimental scheme was proposed for 3$d^{9}$4$s$ in Ni-like HCIs \cite{Ralchenko2017}; however, this has not yet been demonstrated.
It is worth emphasizing that we operate the EBIT with a $low$ magnetic field condition to suppress the Zeeman splitting, and employ the evaporative cooling technique for reduction of the Doppler line broadening, leading to clear observation of the hyperfine-structures.

\section{Result}
\subsection{Concept of the plasma-assisted laser spectroscopy}
In the present experiment, we prepare the (4$d^{9}_{5/2}$5$s$)$_{J=3}$ metastable state of $^{127}$I$^{7+}$ in an EBIT and irradiate a pulse laser to the trapped HCI, as shown in \textbf{Fig.~1~a}.
\textbf{Figure~1~b} shows the experimental scheme used here with the level structure of $^{127}$I$^{7+}$.
Due to the closed 4$d^{10}$ shell structure in the ground state, the excitation energy to the first excited fine-structure level (4$d^{9}_{5/2}$5$s$)$_{J=3}$ is relatively high at approximately 47 eV.
Although such high-energy states are generally de-excited in a short time, the lifetimes for every hyperfine-structure level in (4$d^{9}_{5/2}$5$s$)$_{J=3}$ are longer than 10~s because their de-excitation processes are strongly forbidden, except for $M3$ and hyperfine-mixing $E2$ transitions.
From the long-lived fine-structure level, the extreme ultraviolet (EUV) electric-quadrupole ($E2$) transition ((4$d^{9}_{3/2}$5$s$)$_{J=2}$ $\rightarrow$ (4$d^{10}$)$_{J=0}$)  was induced by pulsed laser excitation $via$ the $M1$ transition ((4$d^{9}_{5/2}$5$s$)$_{J=3}$ $\rightarrow$ (4$d^{9}_{3/2}$5$s$)$_{J=2}$).
Both the initial and excited levels in the laser excitation are metastable states without any $E1$ decay path; thus, the natural width of the transitions is estimated to be $\sim$10$^{-5}$ cm$^{-1}$ from the theoretical 4 $\mu$s lifetime of the upper fine-structure level (4$d^{9}_{3/2}$5$s$)$_{J=2}$.
The intrinsically narrow natural width of the laser transitions contributes to the high-resolution in spectroscopic measurements, revealing the hyperfine-structure.

\begin{figure}[t]
\includegraphics[width=8.5cm]{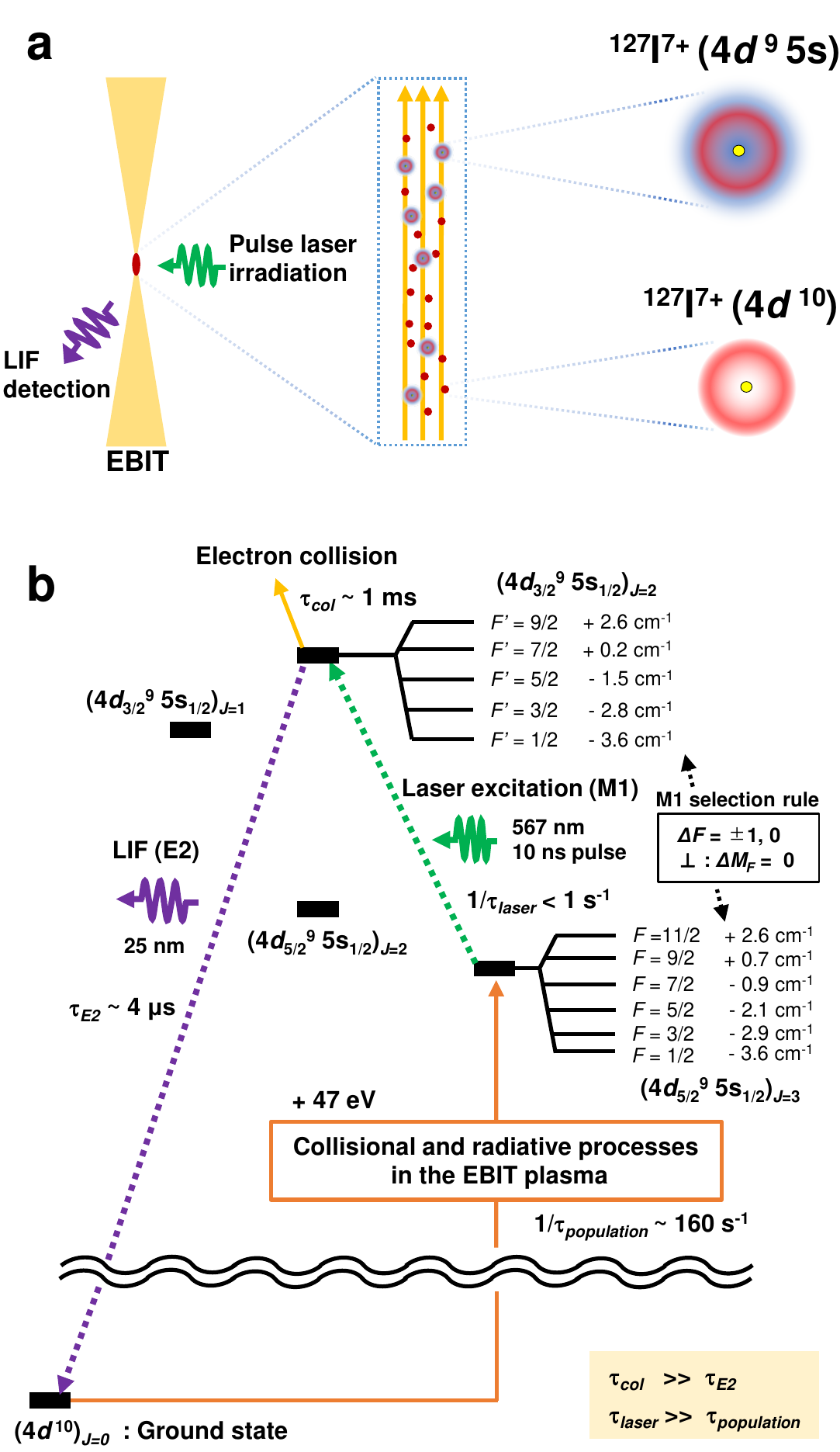}
\caption{\label{Concept} \textbf{Schematic diagram of the present laser spectroscopy.} \textbf{a.} Brief overview of the experiment. \textbf{b.} Experimental scheme with the energy diagram of $^{127}$I$^{7+}$. The population rate to (4$d^{9}_{5/2}$5$s$)$_{J=3}$ was estimated, using collisional-radiative modeling \cite{Kimura2020}. An electron energy of 105 eV and a density of 10$^{10}$ cm$^{-3}$ are assumed in the collisional-radiative rates, corresponding to the present experimental condition. The energy structure and lifetime were calculated using GRASP2018 \cite{GRASP}.}
\end{figure}

\begin{figure*}[t]
\includegraphics[width=17.4cm]{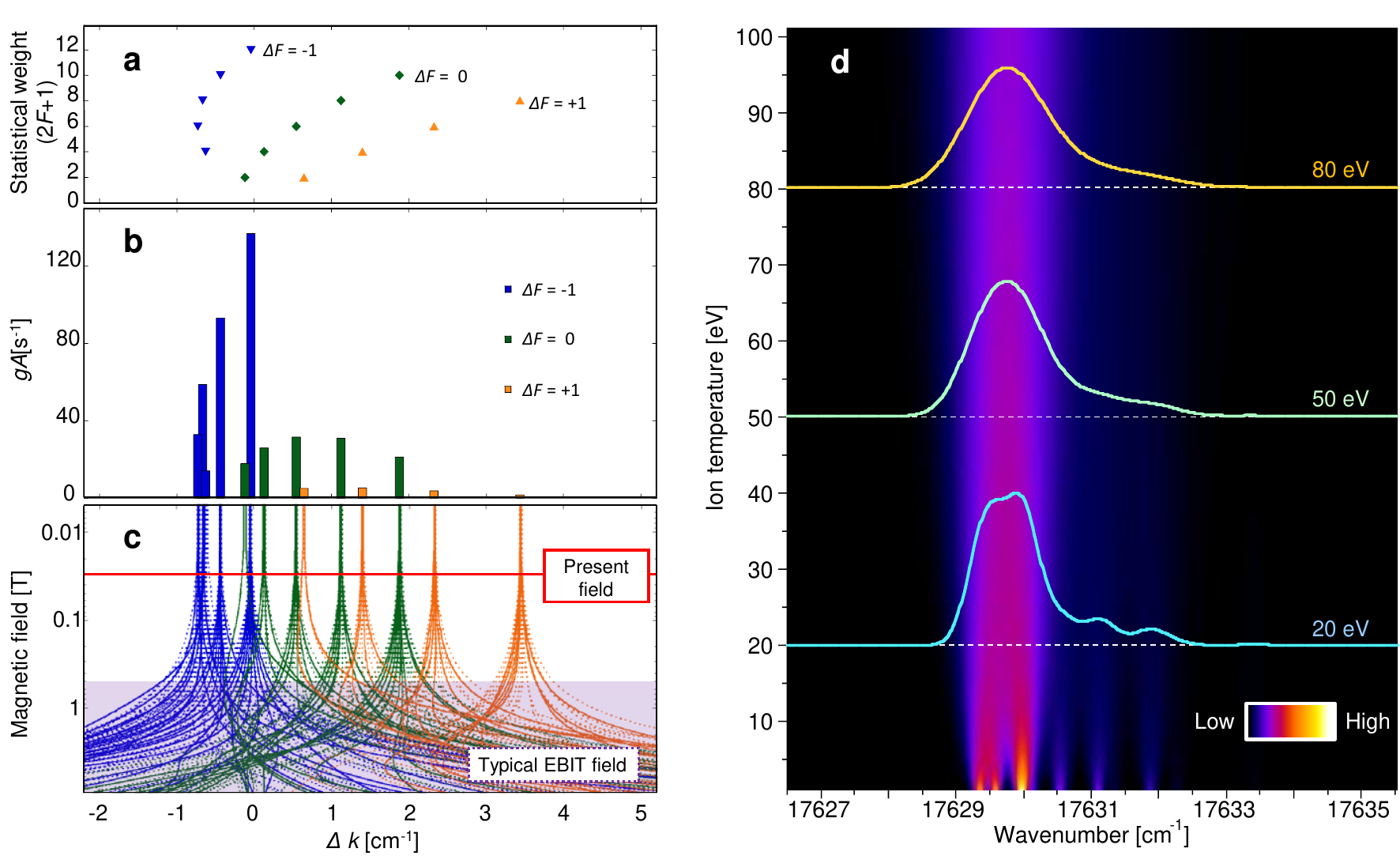}
\caption{\label{EUVSpectrum} \textbf{Theoretical predictions for the hyperfine-structure splitting in the laser excitation.} \textbf{a.} Statistical weights of the initial state (4$d^{9}_{5/2}$5$s$)$_{J=3}$. \textbf{b.} Spectral intensity ($gA$). \textbf{c.} Zeeman splitting. Solid and dotted lines show the $\Delta m_F = 0$ and $\Delta m_F = \pm 1$ transitions, respectively. \textbf{d.} Doppler broadening simulation. The transition energy was taken from the NIST database \cite{NIST}.}
\end{figure*}

The long-lived state (4$d^{9}_{5/2}$5$s$)$_{J=3}$ of $^{127}$I$^{7+}$ was continuously prepared through collisional and radiative processes in the EBIT plasma.
Although increasing the number of trapped ions necessitates an enhanced electron density of the plasma, electron collisions must be suppressed to detect LIF emissions from the laser-excited fine-structure level with a lifetime in the order of microseconds.
We controllably maintained the EBIT electron density below 10$^{10}$ cm$^{-1}$, where the electron collision rate was apporoximately 10$^3$ s$^{-1}$.
The production rate of (4$d^{9}_{5/2}$5$s$)$_{J=3}$ under an electron density of 10$^{10}$ cm$^{-1}$ was estimated to be $\sim$160 s$^{-1}$ from collisional-radiative modeling \cite{Kimura2020}.
Thus, the excitation rate by the pulse laser was operated at less than 1 s$^{-1}$ avoiding the depletion of the population in the initial state (4$d^{9}_{5/2}$5$s$)$_{J=3}$.
This well-designed closed transition cycle, including the plasma processes, maintains the sequential LIF detection, suppressing the unintended population loss for the initial and excited levels in the laser excitation.
We emphasize that the well-defined laboratory plasma and its control with a deep understanding of excitation and de-excitation dynamics through collisional-radiative modeling allow us the LIF detection, even though the ions are in a plasma.

The hyperfine-structure splitting for each fine-structure level is given by the magnetic-dipole hyperfine-structure constant $A_{hfs}$, the electric-quadrupole hyperfine-structure constant $B_{hfs}$, and the quantum numbers $J$, $F$ and $I$.
$A_{hfs}$ and  $B_{hfs}$ are relevant values to the atomic structure and the nuclear moments.
We calculated these constants and the splittings for (4$d^{9}_{5/2}$5$s$)$_{J=3}$ and (4$d^{9}_{3/2}$5$s$)$_{J=2}$ using the atomic calculation code GRASP2018 \cite{GRASP} as shown in \textbf{Fig.~1~b}.
In general, a contracted electron orbital in HCIs enhances the hyperfine-structure splitting. 
Additionally, in the present atomic system, the 5$s$ electron orbital in the HCI has a significant overlap with the nucleus, resulting in the large hyperfine-structure splitting.
In fact, our theoretical calculation shows that the $A_{hfs}$ constant for (4$d^{9}_{5/2}$5$s$)$_{J=3}$ in $^{127}$I$^{7+}$ is six times larger than that for 4$d^{9}_{5/2}$ in $^{127}$I$^{8+}$, even though the binding energy is smaller.

\subsection{Prediction of the spectral profile}
The hyperfine-structures split the laser transition into 14 components, considering the selection rules for $M1$ transitions ($\Delta F=\pm 1, 0$).
Spectral simulations were performed to predict the features of the laser excitation spectrum, including hyperfine-structures.
\textbf{Figure~2~a} shows the statistical weight of the initial state with the wavenumber deviation $\Delta k$ corresponding to the difference from the original transition energy without the hyperfine-structures.
\textbf{Figure~2~b} shows the spectral intensity $gA$, which is the sum of Einstein $A$ coefficients of the $\Delta m_F$=0 transitions for every magnetic sub-level.
We also calculated the Zeeman splitting using the intermediate magnetic field treatment, as shown in \textbf{Fig.~2~c}. 
Equations and atomic codes are summarized in the Methods section.
According to the simulation, the $M1$ transition lines are distributed over several cm$^{-1}$, and the spectral features under low magnetic field conditions have an asymmetric profile because of their specific quantum numbers and the intrinsic difference in the Racah coefficient between these transitions.
At high magnetic fields above 0.5 T, such as that in a typical EBIT, magnetic sub-level resolved transitions spread over a wide $\Delta k$ range, which hinders the experimental assignment of hyperfine components.
To overcome this difficulty,  we employed the low magnetic field operation (0.03 T) of a compact electron beam ion trap (CoBIT) \cite{Nakamura2008} at the University of Electro-Communications.
Although this magnetic field is not classified as the weak field limit, it is sufficiently low enough to suppress the Zeeman splitting below 0.05 cm$^{-1}$.
This quasi Zeeman-free condition enables us to resolve the hyperfine-structures.

In the present experiment, line broadening is expected, mainly caused by the Doppler effect due to ion motion in the EBIT, since the Zeeman splitting, laser linewidth, and natural width of the transition are all less than 0.05 cm$^{-1}$.
\textbf{Figure~2~d} shows the spectral broadening simulation under the assumption of a Maxwellian distribution for the kinetic energies of the trapped ions in the EBIT \cite{Orts2007}.
In the low-temperature region, the width of each transition line becomes narrow, and the associated features originating from the $F=7/2$ $\rightarrow$ $F^{'}=7/2$ and $F=9/2$ $\rightarrow$ $F^{'}=9/2$ transitions appear around 17631-17632 cm$^{-1}$.
The simulation shows that the ions should be cooled below 20 eV to characterize the hyperfine-structures and determine the spectroscopic parameters.

\subsection{Measurement of the laser-induced fluorescence}
\textbf{Figure~3} shows the LIF detection in the present experimental setup.
From CH$_3$I vapor, $^{127}$I$^{7+}$ ions were produced by an electron beam and stored in an electrostatic trap potential formed by the electron beam and three successive cylindrical drift-tube (DT) electrodes \cite{Kimura2020}.
To excite the $M1$ transitions, we used a wavelength-tunable dye laser (Sirah Cobra-Stretch with the dual 3000 lines/mm grating option, Exciton Rhodamine 6G/ Ethanol dye solution) pumped by the second harmonic of an Nd:YAG nanosecond-pulse laser (Cutting Edge Optics Gigashot, 10 ns) with a 100 Hz repetition rate.
The wavelength of the laser was monitored using a high-precision wavemeter (High Finesse, WS-6-600).
The laser pulse energy was tuned and maintained at approximately 7 mJ/pulse by the combination of a $\lambda$/2 wave plate and polarizer.
The polarization of the laser was perpendicular to the magnetic field so that only the $\Delta m_F$ = 0 transitions were excited.
LIF was monitored with a time-resolving extreme ultraviolet (EUV) spectrometer consisting of an aberration-corrected concave grating (Hitachi 001-0660) and a position-sensitive detector (PSD, Quantar Technology Inc., model 3391)  \cite{Tsuda2017}.
A typical $^{127}$I$^{7+}$ emission spectrum obtained with the EUV spectrometer is shown in \textbf{Fig.~3~b}.
Owing to the large difference in wavelength, the LIF signal can be readily distinguished from the laser-scattering noise.
The prominent line at 49 eV is the targeted $E2$ transition ((4$d^{9}_{3/2}$5$s$)$_{J=2}$ $\rightarrow$ (4$d^{10}$)$_{J=0}$), which was used to detect laser excitation.
To distinguish the LIF signal from the continuous emission excited by collisional and radiative processes in the plasma, the time spectrum of the $E2$ transition was recorded with a Multi-Channel-Scaler (MCS) as shown in \textbf{Fig.~3~c}.
The observed lifetime of the LIF signal was approximately 4 $\mu$s, which agrees with our theoretical calculation.

\begin{figure}[t]
\includegraphics[width=8.7cm]{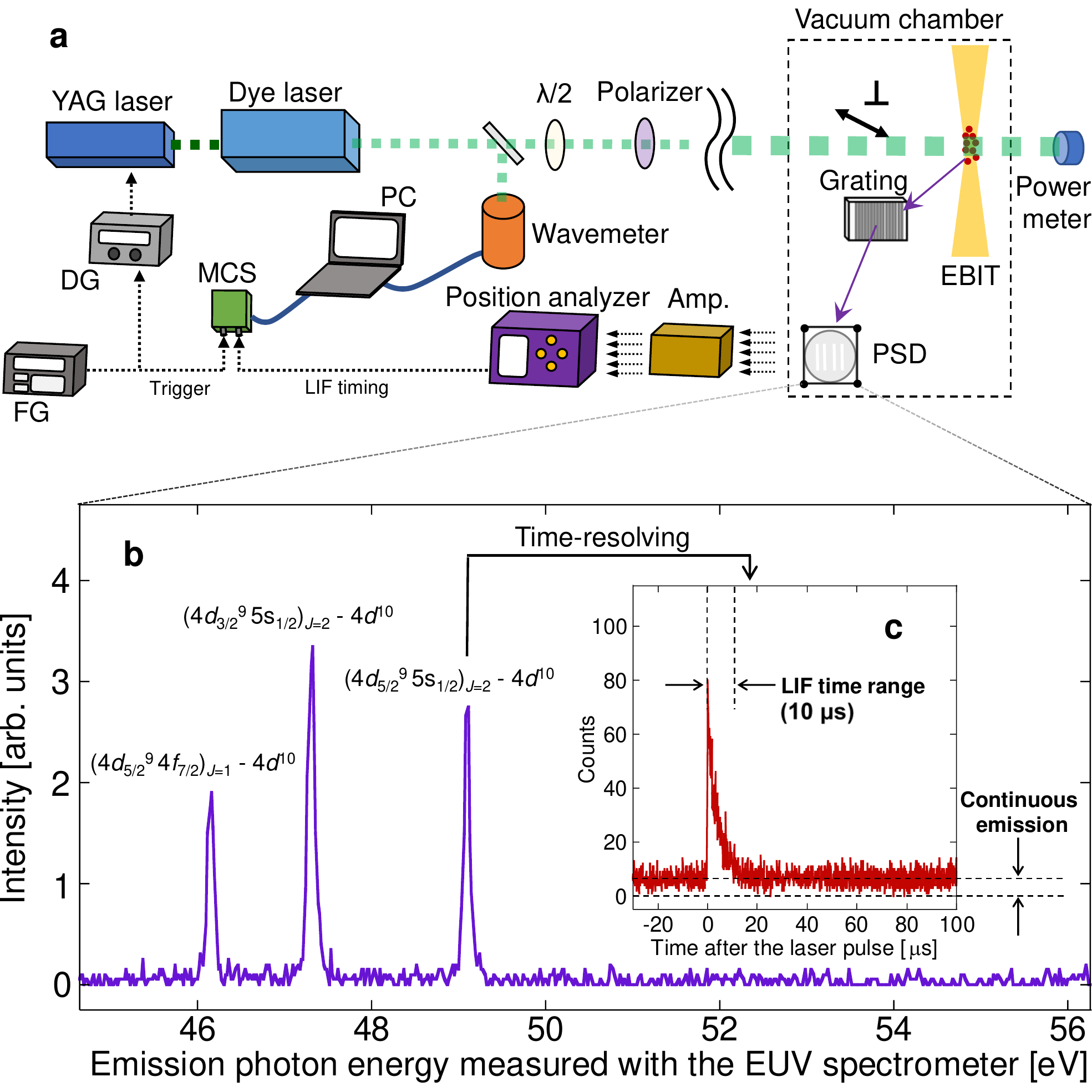}
\caption{\label{EUVSpectrum} \textbf{Overview of the LIF detection setup.} 
\textbf{a.} Schematic of the experimental setup.
\textbf{b.} Typical EUV emission spectrum.
\textbf{c.} Time-resolved signal for the laser-induced $E2$ transition fluorescence.}
\end{figure}

\begin{table*}[t]
\caption{\label{Results} \textbf{Summary of the experimental and theoretical values for (4$d^{9}_{5/2}$5$s$)$_{J=3}$ and (4$d^{9}_{3/2}$5$s$)$_{J=2}$ in $^{127}$I$^{7+}$.} The theoretical results are obtained by Multi-configuration Dirac-Fock calculations combined with the relativistic configuration-interaction (RCI) approach, as described in the Methods section. The RCI corrections of the generalized Breit interaction and the QED effects (self-energy and vacuum polarization) are separately shown.}
\begin{ruledtabular}
\begin{tabular}{cccccc}
 & \multicolumn{1}{c}{Experiment} &\multicolumn{1}{c}{Theory}& \multicolumn{1}{c}{MCDF}&\multicolumn{1}{c}{Breit} &\multicolumn{1}{c}{QED}   \\
\hline
$A_{hfs}$  [GHz]&10.3($\pm$0.3$_{stat}$  $\pm$0.3$_{sys}$)  &10.39($\pm$0.05)& 10.41&  $-$1.7$\times$$10^{-2}$& $+$3.3$\times$$10^{-3}$ \\
$B_{hfs}$  [GHz]& 2.9($\pm$1.8$_{stat}$  $\pm$0.3$_{sys}$) & 2.32($\pm$0.02)& 2.37&  $-$4.3$\times$$10^{-2}$& $+$4.0$\times$$10^{-4}$  \\
$A_{hfs}^{'}$  [GHz]&15.8($\pm$0.3$_{stat}$  $\pm$0.3$_{sys}$) &15.33($\pm$0.03)& 15.45& $-$1.2$\times$$10^{-1}$& $+$7.5$\times$$10^{-3}$   \\
$B_{hfs}^{'}$  [GHz]&1.5($\pm$1.4$_{stat}$  $\pm$0.3$_{sys}$)  &2.02($\pm$0.01)& 2.05&  $-$2.8$\times$$10^{-2}$& $+$3.0$\times$$10^{-4}$   \\
$k_0$  [cm$^{-1}$]&17633.67($\pm$0.02$_{stat}$  $\pm$0.03$_{sys}$) & 17616($\pm$22) & 18016   & $-$418 & $+$18    \\
\end{tabular}
\end{ruledtabular}
\end{table*}

\subsection{LIF spectrum and its analysis}
\textbf{Figure~4} shows the laser wavelength spectrum of the LIF signal.
It took a few hours to accumulate LIF counts at each wavelength.
To compensate for possible fluctuation in the number of stored ions, the LIF counts were normalized by continuous $E2$ counts, which are regarded to be proportional to the ion number. 
The vertical axis in \textbf{Fig.~4} represents the ratio of LIF signal counts within 10~$\mu$s after laser irradiation to the total counts, including the continuous emission within the interval of laser irradiation (10 ms). 
The background level due to continuous plasma emission in the ratio was directly determined to be 0.001 from the time range ratio \mbox{(10~$\mu$s/10~ms)}.
In a preliminary scan, a broad main structure at 17633 - 17634 cm$^{-1}$ and associated structures at 17634 - 17636 cm$^{-1}$ were observed.
We considered that the associated structures were important for understanding the origin of the spectral features.
Therefore, a finer wavelength scan of this region was performed to obtain higher statistics, as shown in \textbf{Fig.~4}.
In the fine scan, all DT electrodes were set to the ground potential to suppress the Doppler line broadening.
While the axial potential formed by the electron beam still enables the storage of $^{127}$I$^{7+}$ ions, the resulting shallow axial potential allows for evaporative cooling, yielding a reduction in the spectral linewidth \cite{Beiersdorfer1996, Orts2007}.
Comparing results to the spectral simulation, we conclude that the main and associated features are mainly composed of ${\Delta F}$ = -1 and 0 transitions, respectively.

\begin{figure}[t]
\includegraphics[width=8.6cm]{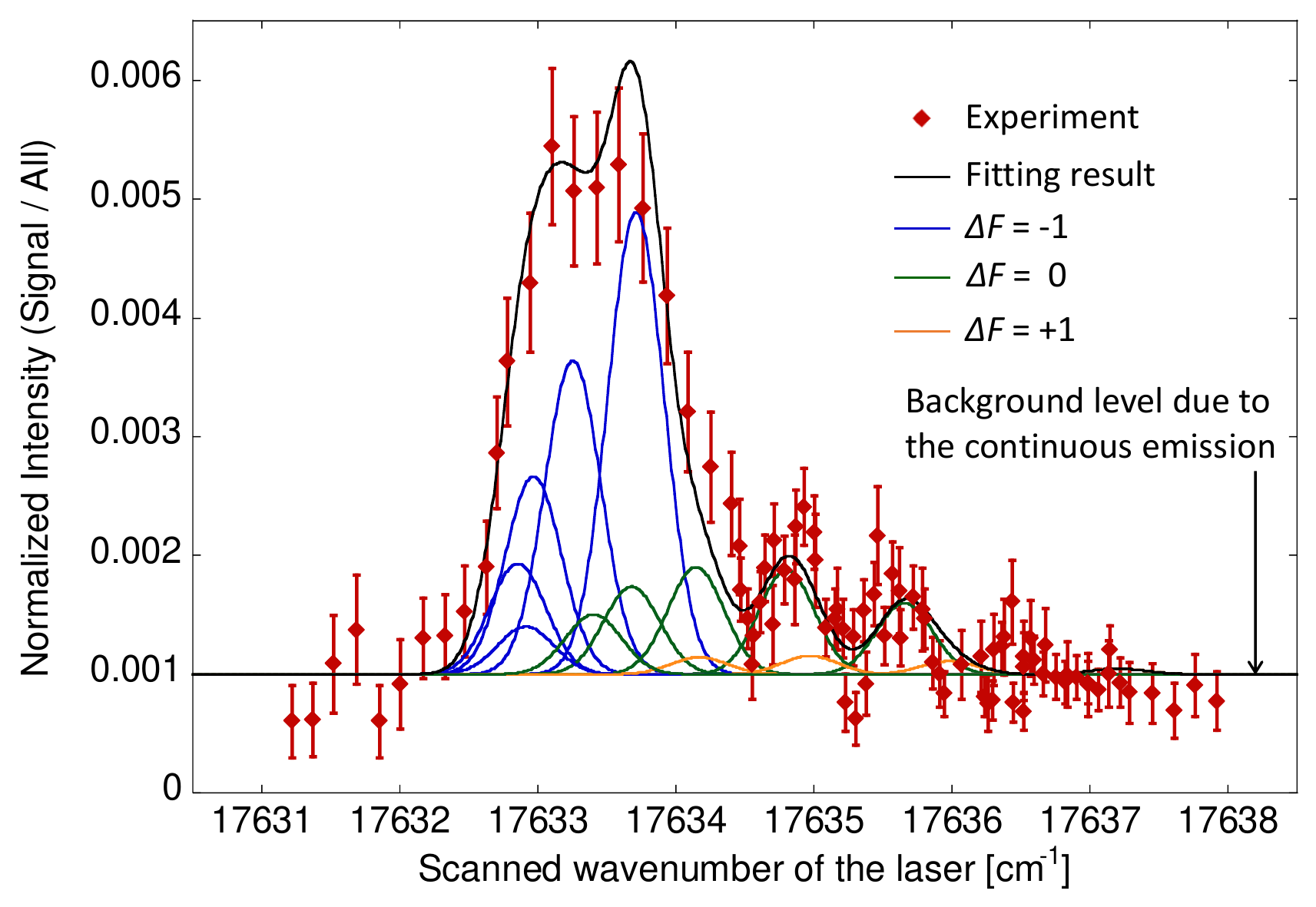}
\caption{\label{LIF spectrum} \textbf{Experimental LIF spectrum.} The fitting result  is shown with separated profiles for every transition. The error bar for each data point represents the statistical error.}
\end{figure}

To determine the hyperfine-structure constants, the observed spectrum was fitted with a model function incorporating all hyperfine-resolved transitions with the same Doppler width $k_D$.
The model function is described in the Methods section.
The center of the transition energy for each line is determined by the original transition energy $k_{0}$ for the fine-structure transition and the hyperfine-structure constants for each of the initial and excited fine-structure levels.
The relative line intensity for each transition was defined by the simulation intensity in \textbf{Fig.~2~b}.
Consequently, the seven parameters $A_{hfs}$, $B_{hfs}$, $A_{hfs}^{'}$, $B_{hfs}^{'}$, $k_0$, $k_D$, and total spectral intensity $I_0$ were determined in the fitting procedure,
where the primed quantities refer to the excited fine-structure level.
Note that theoretically calculated $A$ coefficients were used to determine the simulated intensities.
The intensity model mainly depends on the Racah coefficients whereas the actual intensity is slightly different because of the hyperfine-structure mixing with other fine-structure levels.
We theoretically evaluated the hyperfine-mixing effect on the transition intensities and found that this contribution is less than 0.05 $s^{-1}$ for every $A$ coefficient.
This ambiguity in the model function affects the determination of the experimental values by less than 1$\%$ of the experimental uncertainty and is thus considered not significant.
The statistical population in the intensity model may also be disturbed by the lifetime difference between hyperfine levels \cite{Trabert2007}.
However, in the case of $^{127}$I$^{7+}$, it is not necessary to consider such possibilities because the radiative decay rates are at least $10^3$ times longer than the collisional-radiative rate supplying the population to the metastable fine-structure level;  hence, the populations are not quenched by spontaneous decay to the ground state.
Thus, we consider the intensity model used here to be reasonable for determining the constants.
The fitted profiles for each hyperfine component and their convolution are shown in Fig.~4.
The obtained parameters are listed in \textbf{Table~I} along with the theoretical values calculated by the GRASP code taking the Breit and quantum electrodynamic (QED) effects into account.
The reduced-chi-square in the fitting was 1.1.
From the laser wavelength stability during the measurement, the systematical errors for the hyperfine-structure constants were estimated to be 0.3 GHz.
The systematic error of $k_0$ is 0.03 cm$^{-1}$ taking the absolute accuracy of the wavemeter into account.
The resulting $k_D$ is 0.46 ($\pm$ 0.04) cm$^{-1}$ corresponding to an ion temperature of 15 ($\pm$ 2)  eV under the assumption that Doppler broadening dominates in the spectral line width.
The shallow trapping potential formed by the zero DT voltage and the thin electron density with a low magnetic field enabled the trapped ions to be evaporatively cooled down to the low temperature similar to that found in forced evaporative cooling by turning off the electron beam \cite{Mackel2011}.
The narrow Doppler width realized in the present measurement is essentially important for resolving the hyperfine splitting.

\begin{figure*}[t]
\includegraphics[width=14cm]{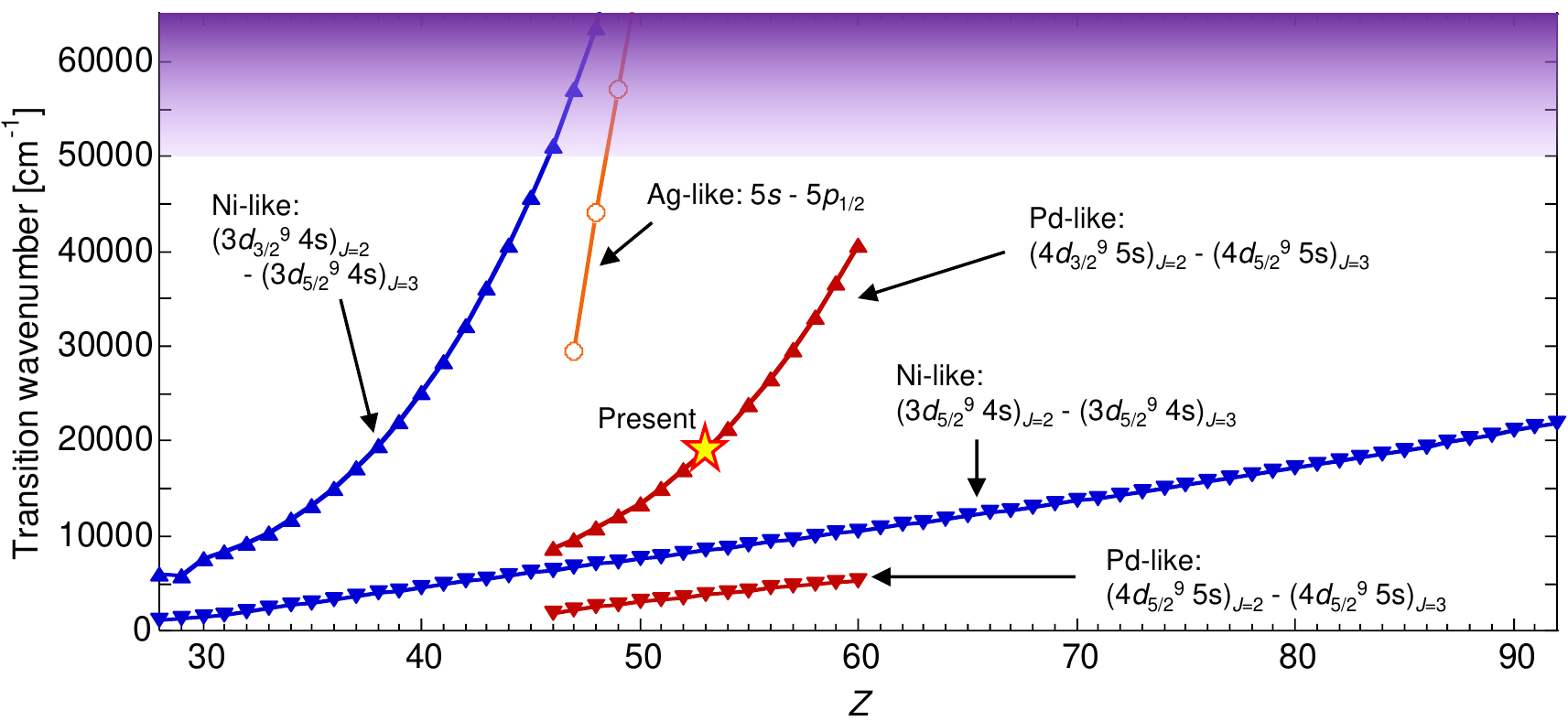}
\caption{\label{Z-Dependence} \textbf{Potential of the present laser spectroscopy for studying heavy nuclei.} Atomic number ($Z$) dependence of the theoretical transition energies in Pd-like and Ni-like ions along with the transition energies of the $E1$ transition 5$s$ - 5$p_{1/2}$ in Ag-like ions is shown for comparison. The purple shaded region indicates wavelength ranges where it is generally difficult for lasers to reach ($<$ 200 nm).}
\end{figure*}

Because the nuclear spin and moments of $^{127}$I are well-known \cite{Moment_Table}, the experimentally obtained hyperfine-structure constants are used to validate the theoretical calculation of the electron orbital for each fine-structure level.
Experimental results obtained here show reasonable agreement with theoretical values, considering uncertainties. 
The transition energy $k_0$ obtained by the experiment agrees with the theoretical calculation within approximately 0.1 $\%$.
The Breit interaction shifts the transition energy by 2 $\%$, which is important to reproduce the experimental result.
The energy of each fine-structure level has been compiled in the NIST database \cite{NIST} from several experimental results using passive spectroscopy, and the uncertainty was evaluated to be 100 cm$^{-1}$.
In the present study, we clearly demonstrate that our active spectroscopic approach succeeded in directly measuring the transition energy and greatly improving its uncertainty.
This clarified the importance of the relativistic configuration-interaction (RCI) correction for the Breit interaction.

\section{Discussion}
We have demonstrated laser spectroscopy of forbidden transitions between metastable states of HCIs stored in an EBIT by employing Pd-like $^{127}$I$^{7+}$.
The laser excitation spectrum of the HCIs in a quasi Zeeman-free low magnetic field revealed distinct hyperfine-structures, which provided evidence for the enhancement of the hyperfine interaction via a contracted electron cloud with a 5$s$ valence.
The resulting hyperfine-structure constants of highly charged ions with a $well$-$known$ nucleus provide a benchmark for atomic orbital calculations with relativistic many-electron correlations, allowing for discussion of the electron cloud in the vicinity of the nucleus for each fine-structure level.
Even though the transition observed in this study is not a proposed HCI clock candidate, the building of a benchmark to understand hyperfine-structures in many-electron HCIs makes a significant step toward developing such clocks, especially that the most promising current clock candidates with a 5$s$-4$f$ level crossing \cite{Berengut2010, Berengut2011, Berengut2012-1, Dzuba2012, Safronova2014-1, Safronova2014-2, Safronova2014-3, Dzuba2015, Windberger2015,  Dzuba2016, Nandy2016, Kozlov2018} are intrinsically accompanied by a similar characteristic electron configuration having a single 5$s$ electron.
The established benchmarks would serve as reference to enhance the understanding of the hyperfine interaction in many-electron HCIs.

The present hyperfine-structure-resolved laser spectroscopy offers future possibilities for experimental studies in nuclear physics using HCIs, such as a broad investigation of $unknown$ nuclear spins and moments and hyperfine anomalies with a specialized EBIT for highly charged radionuclide ions \cite{Vondrasek2016}.
Recently, high-precision atomic orbital calculations have been extensively developed to investigate nuclear properties from hyperfine-structure constants \cite{Campbell2011, Safronova2013, Thielking2018, Li2021, Porsev2021}.
While this approach had the privilege for particular nuclei with optically accessible transitions at low valence ions, the extension of the target to HCIs enables the selection of an atomic-level structure by changing the charge state.
In addition, an $ns$-valence state enhancing its hyperfine interaction can be arbitrarily selected, in contrast to neutral atoms or single charged ions whose electron configurations are fixed.
We consider that the present LIF scheme has the potential to be a versatile method for investigating heavy nuclear properties because it can be directly applied to the three transitions in Pd-like ions and Ni-like ions, as follows.
Pd-like ions have an additional excitation pathway from the long-lived (4$d^{9}_{5/2}$5$s$)$_{J=3}$ state to the (4$d^{9}_{5/2}$5$s$)$_{J=2}$ excited fine-structure level via an $M1$-allowed transition.
Ni-like ions have a similar energy level system with the 3$d^{10}$ closed ground state.
Therefore, the investigation of the (3$d^{9}_{5/2}$4$s$)$_{J=3}$ $\rightarrow$ (3$d^{9}_{3/2}$4$s$)$_{J=2}$ and (3$d^{9}_{5/2}$4$s$)$_{J=3}$ $\rightarrow$ (3$d^{9}_{5/2}$4$s$)$_{J=2}$ transitions with the present laser spectroscopy scheme is also possible.
\textbf{Figure 5} shows the atomic number dependence of the transition energies theoretically calculated by the FAC code \cite{Gu2008}, along with optical $E1$ transition energies of Ag-like ions for comparison.
In the $Z$$>$60 region, Pd-like ions are not shown because they are not suitable for the present laser spectroscopy since the (4$d^{9}_{5/2}$5$s$)$_{J=3}$ level is not metastable due to the 4$d^{9}$5$s$ - 4$d^{9}$4$f$ level crossing \cite{Ivanova2009}.
In contrast to the typical electronic $E1$ transitions (e. g., $5s$  - $5p_{1/2}$ in Ag-like ions), the present $M1$ transitions with a number of inner-shell electrons show a gradual increment in the transition energy with an increasing atomic number.
As a result, in most regions where $Z$$>$30, transitions in either Pd- or Ni-like ions are available for laser excitation.

The present demonstration achieved precise determination of the transition energy with an uncertainty of $\Delta k$/$k_0$ $\simeq$ 3$\times$$10^{-6}$ for the $M1$ transition, which is inaccessible by emission spectroscopy due to the rapid $E2$ decay path from the upper fine-structure level.
From a technical perspective, we have succeeded in extending experimental targets in precision spectroscopy for trapped HCIs from spontaneously emitted transitions to non-luminescent transitions, providing a wider variety of spectroscopic targets.
This offers new opportunities, including the investigation of HCI clock candidates.
Prospects of the new type of atomic clock have also spurred developments in HCI laser spectroscopy with part-per-million uncertainties.
Several laser spectroscopic methods for forbidden transitions of trapped HCIs free from a systematic Doppler shift have been reported \cite{Mackel2011, Egl2019}.
These demonstrations employed a simple fine-structure transition $^2P_{1/2}$-$^2P_{3/2}$ in the doublet ground term of Ar$^{13+}$, whose transition energy is precisely known from direct wavelength measurements using passive spectroscopy \cite{Dragnic2003, SoriaOrts2006}, leading to recent quantum logic spectroscopy \cite{Micke2020, King2022}.
However, many potential candidates for the HCI atomic clock are not directly accessible using such passive spectroscopy because they possess complicated level structures with hyperfine splitting, and their clock transitions are not emitted in plasmas owing to their long lifetime.
Therefore, the extension of laser spectroscopy of HCIs to complex systems is a new challenge to overcome.
The present scheme can be applied to investigate the repump transitions and measure the hyperfine-structure constants of a clock state; otherwise is not accessible using passive spectroscopic methods.

Finally, we discuss the possibilities of other applications using the present spectroscopic technique.
The time-resolved LIF detection technique can be applied to hyperfine-level selective lifetime measurements without cascade contributions from higher levels, leading to deep investigations of hyperfine interaction \cite{Trabert2007} including hyperfine-induced-interference effects \cite{Yao2006} which have not yet been experimentally observed.
The demonstration of laser spectroscopy of forbidden transitions for HCIs in a laboratory plasma is expected to trigger the development of active sensing to investigate plasma conditions such as the plasma ion temperature and magnetic field, contributing to plasma diagnostics.

\section{Methods}
\subsection{Theoretical transition energy}
The transition energy between the fine-structure levels in 4$d^{9}$5$s$ is obtained in the framework of the multi-configuration Dirac-Fock (MCDF) calculation, combined with the relativistic configuration-interaction (RCI) approach, using the atomic structure calculation code package GRASP2018 \cite{GRASP}. 
First, MCDF calculations were performed with an active space of single- and double-electron excitation from the 4\textit{s}, 4\textit{p}, and 4\textit{d} orbitals up to 8\textit{g}. 
Each \textit{nl} orbital (\textit{n}=5-8, \textit{l}=\textit{s,p,d,f,g}) is added individually, with each outer orbital optimized while keeping all inner orbitals fixed.
Furthermore, to account for core-core and core-valence correlations with the inner orbitals, excitations from the  3$l$ ($l = s, p, d$)  orbitals were also included.
This active space treatment led to 3,300,000 $jj$-coupled configurations. 
Moreover, corrections from the Breit interaction, that is transverse photon interactions in the low-frequency limit and QED corrections separating the self-energy and vacuum polarization corrections, are included in the RCI calculation.

\subsection{Theoretical hyperfine-structure constants}
Using the GRASP2018 code, we calculated the magnetic-dipole hyperfine-structure constant $A_{hfs}$ and electric-quadrupole hyperfine-structure constant $B_{hfs}$ for each fine-structure level ((4$d^{9}_{5/2}$5$s$)$_{J=3}$ and (4$d^{9}_{3/2}$5$s$)$_{J=2}$).
For the input values, the nuclear magnetic moment $\mu_I$ and nuclear electric quadrupole moment $Q_I$ were taken from \cite{Moment_Table}.
GRASP2018 also provides hyperfine-structure splitting calculations taking hyperfine-mixing with other fine-structure levels into account.
We compared two calculation cases, i.e. with and without considering shifts due to hyperfine-mixing.
It has been found that the resulting shift in energy for each hyperfine-structure level would be unobservable in the experiment.
Each correction for the hyperfine-structure constants was also evaluated using the same procedure as for the transition energy calculation.

\subsection{Estimation of uncertainties for the theoretical values}
There are several origins of uncertainties on the theoretically calculated values due to various corrections.
The uncertainties in the transition energy were estimated as follows:
The present MCDF-CI calculations involve valence-valence(VV), core-valence(CV), and core-core(CC) correlations.
They are considered by including configurations generated from all single- and double-excitations from the 3$s$, 3$p$, 3$d$, 4$s$, 4$p$, 4$d$, and 5$s$ orbitals to the active spaces (AS) of virtual orbitals.
Here we define four virtual orbital sets as follows: AS1 = \{$5p$, $5d$, $5f$, $5g$\}, AS2 = AS1 + \{$6s$, $6p$, $6d$, $6f$, $6g$, $6h$\}, AS3 = AS2 + \{$7s$, $7p$, $7d$, $7f$, $7g$, $7h$\} and, AS4 = AS3 + \{$8s$, $8p$, $8d$, $8f$, $8g$, $8h$\}. 
We estimated the convergence with the size of the basis set by taking the difference between the transition energies calculated in the AS3 and AS4 stages ($\simeq$ 22 cm$^{-1}$).
We adopted this value for the uncertainty of the electron correlation calculation because additional correlations were not expected to exceed this convergence value even when more virtual orbitals we considered.
Another possible source of uncertainty is related to the calculated contribution from the Breit interaction. 
The corresponding correction value was perturbatively computed in the RCI calculation through the exchange of a single transverse photon in the low-frequency limit \cite{GRASP, Chantler2014}.
Comparing the Breit contributions obtained from the perturbative approach to calculations in which the Breit term is included in a variational SCF process allows us to estimate the uncertainty due to the perturbative approach \cite{Koziol2020, Safronova2018}.
Previous studies have found that the magnitude of this effect is less than 0.3 ${\%}$ in the Ni-like case (among others) \cite{Koziol2020, Safronova2018} and that it is significantly reduced when the active space is expanded.
Here, we assume the maximum error owing to the perturbative treatment and adopt 0.5 ${\%}$ of uncertainty related to the Breit correction.
The resulting uncertainty is 2 cm$^{-1}$.
The QED contributions were smaller than the uncertainties related to correlation effects.
Thus, the resulting uncertainty associated with these terms was negligible.
The uncertainty in the Dirac-Fock term is caused by the uncertainty in the root-mean-square of the nuclear radius, which is insignificant for $^{127}$I  ($\simeq$ 0.008 fm \cite{Size_Table}).
The total uncertainty was calculated by applying the error propagation rule (quadratic summation) to individual sources of uncertainty. 
This yielded an uncertainty of the theoretical transition energy of 22 cm$^{-1}$.
This uncertainty is strongly dominated by the electron correlation.
We also estimated the uncertainty for each $A_{hfs}$ and $B_{hfs}$ constant by employing the same procedure adopted for the transition energy.

\subsection{Simulation of the Zeeman splitting}
The Zeeman splitting for each hyperfine-structure level was calculated using the HFSZEEMAN95 package \cite{HFSZEEMAN,HFSZEEMAN95}, which is based on inputs from GRASP2018 \cite{GRASP}.
This calculation employs an accurate treatment of the intermediate field regime; therefore, the calculated splittings are reliable for a wide range of magnetic fields. 
\textbf{Fig.~2~c} shows the differences in the Zeeman splitting between the initial and excited levels of the laser transition.

\subsection{Simulation of the transition intensities}
The hyperfine-structure-resolved Einstein $A$ coefficients were obtained by the summation of the Einstein $A$ coefficients for every $\Delta m_F = 0$ transition at each initial hyperfine-structure level.
The transition rate for an $M1$ transition between magnetic hyperfine-structure sub-levels $\Gamma^\prime m_F^\prime$ and $\Gamma m_F$ is given by
\begin{eqnarray}
\begin{multlined}
A(\Gamma^\prime m_F^\prime \rightarrow \Gamma m_F)\\
=\frac{2.69735\times10^{13}}{\lambda^3}\sum_{q}|\bra{\Gamma m_F}M^{(1)}_q\ket{\Gamma^\prime m_F^\prime}|^2,
\end{multlined}
\end{eqnarray}
where $M^{(1)}$ is the magnetic dipole operator, and $\lambda$ is the transition wavelength (in \AA).
Using the wavefunction expansion of the magnetic hyperfine substates,
\begin{equation}
|\Gamma m_F> =\sum_{\gamma,J,F}d_{\gamma,J,F}|\gamma I J F m_F>
\end{equation}
together with the Wigner-Eckart theorem (the dipole operator only acts on the electronic space, so we may decouple the nuclear and electronic parts in the reduced matrix element), equation (1) can be re-written as \cite{HFSZEEMAN95}
\begin{eqnarray}
\begin{multlined}
A(\Gamma^\prime m_F^\prime \rightarrow \Gamma m_F)=\\
\frac{2.69735\times10^{13}}{\lambda^3}\sum_{q}|\sum_{\gamma J F}\sum_{\gamma^\prime J^\prime F^\prime}d_{\gamma J F}d_{\gamma^\prime J^\prime F^\prime}\\
\sqrt{(2F^\prime+1)(2F+1)} (-1)^{F-m_F} \\
\begin{pmatrix}
F & 1 & F^\prime\\
-m_F & q & m_F^\prime
\end{pmatrix}
(-1)^{I+J^\prime+F+1}\\
\begin{Bmatrix}
J & F & I\\
F^\prime & J^\prime & 1
\end{Bmatrix}\bra{\gamma J}|M^{(1)}|\ket{\gamma^\prime J^\prime}|^2.
\end{multlined}
\end{eqnarray}
Here, $d_{\gamma J F}$ are the expansion coefficients, and $\bra{\gamma J}|M^{(1)}|\ket{\gamma^\prime J^\prime}$ is the reduced transition matrix element between the fine-structure levels. 
The $A$ coefficients for all $M1$ transitions were calculated using the HFSZEEMAN95 package \cite{HFSZEEMAN,HFSZEEMAN95}.
Finally, we sum the Einstein $A$ coefficients for the $\Delta m_F = 0$ transitions of every $m_F$ sub-level and obtained the transition intensities $gA$ of the $M1$ transitions, as shown in \textbf{Fig.~2~b}.

\subsection{Fitting model}
We determined the hyperfine-structure constants $A_{hfs}$, $B_{hfs}$, $A_{hfs}^{'}$, $B_{hfs}^{'}$ and the fine-structure transition wavenumber $k_0$ using the following model equation to fit experimental data.
\begin{eqnarray}
\begin{multlined}
f(k) =\\
I_{0} \sum_{|F^{'}-F|\leq 1} gA_{F^{'}F} \exp(\frac{4\ln2(k-(k_{0}+k_{F^{'}F}))}{k_D^{2}}). 
\end{multlined}
\end{eqnarray}
Here, it is assumed that each line profile is defined by a Gaussian function with the same line width because the hyperfine-structure splitting energies are significantly smaller than the fine-structure transition energy.
$k_D$ represents the Full-Width at Half Maximum (FWHM) of the transition line profiles.
For $gA_{F^{'}F}$, we employed theoretically calculated results for the relative transition intensities, as described in the previous section.
$k_0$ is the original transition energy between the fine-structure levels ((4$d^{9}_{5/2}$5$s$)$_{J=3}$ $\rightarrow$ (4$d^{9}_{3/2}$5$s$)$_{J=2}$).
$k_{F^{'}F}$ is the hyperfine-structure shift for each transition given by
\begin{eqnarray}
k_{F^{'}F} &=& k_{F^{'}} - k_{F}.
\end{eqnarray}
$k_F^{'}$ and $k_F$ are hyperfine-structure shifts from the original level for (4$d^{9}_{3/2}$5$s$)$_{J=2}$ and (4$d^{9}_{5/2}$5$s$)$_{J=3}$, respectively.
The hyperfine-structure splitting at each fine-structure level is given by 
\begin{eqnarray}
\begin{multlined}
k_{F^{'}} =\\
 \frac{1}{2} A_{hfs}^{'} C^{'} + B_{hfs}^{'} \frac{\frac{3}{4}C^{'}(C^{'}+1)-I(I+1)J^{'}(J^{'}+1)}{2I(2I-1)J^{'}(2J^{'}-1)}
\end{multlined}
\end{eqnarray}
and
\begin{eqnarray}
\begin{multlined}
k_{F} =\\
  \frac{1}{2} A_{hfs} C + B_{hfs} \frac{\frac{3}{4}C(C+1)-I(I+1)J(J+1)}{2I(2I-1)J(2J-1)}.
  \end{multlined}
\end{eqnarray}
Here, $C^{'}$ and $C$ are given by:
\begin{eqnarray}
C^{'} &=& F^{'}(F^{'}+1) - J^{'}(J^{'}+1) - I(I+1) 
\end{eqnarray}
and
\begin{eqnarray}
C &=& F(F+1) - J(J+1) - I(I+1).
\end{eqnarray}
Under the assumption of a Maxwellian distribution, the equation for the ion-temperature and spectral line width is given by:
\begin{eqnarray}
T &=& \frac{Mc^{2}}{8k_{B}ln2}(\frac{k_{D}}{k_{0}})^{2},
\end{eqnarray}
where  $M$, $c$, and $k_B$ are the particle mass, speed of light, and Boltzmann constant, respectively.
Equation (10) can be re-written as:
\begin{eqnarray}
k_{D} &=& \sqrt{\frac{8Tk_{B}ln2}{M}}\frac{k_{0}}{c}.
\end{eqnarray}
Equations (11) and (4) provide theoretical spectra.

\section{Data availability}
The data that support the findings of this study are available from the corresponding author upon reasonable request.

\section{Code availability}
The FAC code is available at https://www-amdis.iaea.org/FAC/, the GRASP2018 code at https://www-amdis.iaea.org/GRASP2K/, and the HFSZEEMAN95 code at http://dx.doi.org/10.17632/rv2vycs7pg.1.

\section{Acknowledgements}
We would like to thank Dr. Haruka Tanji-Suzuki for providing an optical component.
We also thank Dr. Michiharu Wada, Dr. Kunihiro Okada, Dr. Takayuki Yamaguchi, and Dr. Kiattichart Chartkunchand for fruitful discussions.
This work was supported by the RIKEN Pioneering Projects and by the JSPS KAKENHI Grant No.~16H04028, No. 20K20110, and 22K13990.

\section{Author contributions}
N. K. conceived and initiated this work.
N. K., S. K., T. A., and N. Nak. designed the experimental scheme and selected the target ion.
N. K. and Pri. carried out the LIF measurement using a compact EBIT device developed by N. Nak.
Y. K, P. Pip., K. S., and N. Num. provided the support for the EBIT operation.
The laser system was constructed by N. K. and S. K.
The experimental data were analyzed by N. K. and Pri.
The theoretical calculation was performed by Pri.
N. K. prepared the initial manuscript assisted by Pri.,  S. K., T. A., and N. Nak.
All authors discussed the result and reviewed the manuscript.

\end{document}